# Chapter 1

# HYDRODYNAMIC INSTABILITIES IN CLOSE BINARY SYSTEMS

FREDERIC A. RASIO
*Massachusetts Institute of Technology*
*Department of Physics, MIT 6-201, Cambridge, MA 02139, USA*
rasio@mit.edu

**Abstract**

Equilibrium fluid configurations for close binary systems can become *globally unstable*. Instabilities arise from the strong tidal interaction between the two components, which tends to make the effective two-body potential governing the orbital motion steeper than $1/r$. As a result, a circular orbit can become unstable to small radial perturbations. The instability can be either secular or dynamical. In both cases it leads to the coalescence and merging of the two components on a timescale generally much shorter than the lifetime of a stable system. In a secularly unstable system, orbital decay is driven by viscous dissipation and proceeds on a timescale comparable to the tidal circularization time of a stable binary. In a dynamically unstable system, the two stars suddenly plunge toward each other and merge hydrodynamically in just a few orbital periods. These instabilities are relevant to a wide variety of astrophysical problems involving coalescing binaries, such as the formation of blue stragglers, Type Ia supernovae, the orbital decay of massive X-ray binaries, and the calculation of gravitational radiation signals from close neutron star binaries.

## 1.1 Introduction

### 1.1.1 The Stability of Self-Gravitating Fluid Equilibria

Perhaps the most fundamental question that one can ask about a binary star system is whether the equilibrium configuration of the fluid in the system is a stable one. For well-separated components, the two stars behave like point masses and the dynamics of the system is purely Keplerian. In that case, of course, we do not expect any instabilities to occur. For very close components, however, one must treat the binary as a self-gravitating fluid system in three dimensions. In this case, two types of *global fluid instabilities* can occur in general. A *dynamical instability* conserves total energy and grows on the dynamical timescale $t_{\rm dyn} \sim a^{3/2}/(GM)^{1/2}$, where $a$ is the binary separation and $M$ is the stellar mass. A *secular instability* grows only in the presence of a dissipation mechanism and allows the system to decrease its total energy on a timescale $t_{\rm diss} \sim a^2/\nu$, where $\nu$ is the effective viscosity of the fluid.

Well-known examples of such secular and dynamical fluid instabilities are those affecting rapidly rotating single stars. Recall that along a sequence of equilibrium configurations for a single rotating star, one encounters a secular instability at $T/|W| \simeq 0.14$ and, further along the sequence, a dynamical instability at $T/|W| \simeq 0.27$ (see, e.g., Shapiro and Teukolsky 1983, Chap. 7). Here $T/|W|$ is the ratio of kinetic energy of rotation to gravitational binding energy. The analogous dimensionless ratio for a close binary system is simply $R/a$, the ratio of stellar radius to binary separation, which is a direct measure of the strength of the tidal interaction between the two components, i.e., a direct measure of the deviation from point-mass Keplerian behavior. As $R/a$ increases along a sequence of equilibrium binary configurations with progressively closer components, one can encounter a secular instability first, and later possibly also a dynamical instability. Of course these instabilities cannot always be reached along a particular equilibrium sequence. This is because the equilibrium sequence may terminate before even reaching the secular stability limit. For single rotating stars, the equilibrium sequence terminates at the onset of equatorial mass shedding, which can occur for a value of $T/|W|$ smaller than the secular stability limit at $T/|W| \simeq 0.14$. This is often true for sequences of *uniformly rotating* configurations (cf. Bodenheimer and Ostriker 1973). Similarly, for close binaries, the equilibrium sequence may terminate at the Roche limit (or, for contact configurations, at the onset of mass shedding through the *outer* Lagrangian point) before encountering an instability. Whether unstable equilibria exist depends mainly on the equation of state of the fluid. For a sufficiently compressible fluid one finds that all equilibrium configurations are stable. This is simply because stars with very centrally condensed mass profiles are not much affected by the tidal interaction and continue to behave dynamically as point masses even in very close binaries. For a sufficiently incompressible fluid, however, both secular and dynamical instabilities can occur.

Instabilities (both secular and dynamical) of close binary configurations have been studied previously by Chandrasekhar and collaborators (see Chandrasekhar 1969 and references therein) in the context of classical equilibrium solutions for incompressible fluids. The secular instabilities have been discussed in the past mainly in the context of classical tidal dissipation theory (Alexander 1973; Counselman 1973; Zahn 1977; Hut 1980, 1981, 1982). Recently, a more unified and general treatment of these instabilities (including also those of single rotating stars) has been developed by the author and his collaborators. Rasio and Shapiro (1992, 1994, 1995, hereafter RS1–3 or collectively RS) have studied numerically the stability limits for systems containing polytropic components, and have followed the coalescence of unstable systems using numerical hydrodynamics calculations in three dimensions. At the same time, a comprehensive analytic study of the stability properties of close binaries containing polytropes, using an energy variational method, was undertaken by Lai, Rasio, and Shapiro (1993a,b; 1994a,b,c, hereafter LRS1–5 or collectively LRS).

### 1.1.2 Astrophysical Motivation

When a binary system becomes unstable, rapid orbital decay is inevitable, even in the absence of any angular momentum loss mechanism. The orbital angular momentum is simply transferred to the fluid as the instability grows, and the total angular momentum of the system remains constant. The orbital decay of a *dynamically* unstable system proceeds on the dynamical time $t_{\rm dyn}$, and the two stars typically undergo complete coalescence in just a few rotation periods. Instead, for a *secularly* unstable system, the orbital decay is driven by internal viscous dissipation in the fluid and it proceeds on the (usually much longer) dissipation timescale $t_{\rm diss}$. This timescale is essentially the same as the timescale for tidal circularization of a stable system (see, e.g., Zahn 1977). Although their growth time is usually much longer than for dynamical instabilities, secular instabilities also tend to drive the binary to coalescence.

Binary coalescence has been associated with a number of astrophysical phenomena of great current interest, which we now review briefly.

#### 1.1.2.1 Coalescing Neutron-Star Binaries

Coalescing neutron-star binaries are most important sources of gravitational radiation in the Universe, and are the primary targets for LIGO, the Laser-Interferometer Gravitational-wave Observatory, now under construction in the United States (Abramovici et al. 1992). Statistical arguments based on the observed local population of binary pulsars with probable neutron star companions lead to an estimate of the rate of neutron star binary coalescence in the Universe of order $10^{-7}\,{\rm yr}^{-1}\,{\rm Mpc}^{-3}$ (Narayan, Piran and Shemi 1991; Phinney 1991). Finn and Chernoff (1993) estimate that an advanced LIGO detector could observe about 70 events per year. In addition to providing a major confirmation of Einstein's theory of general relativity, the detection of gravitational waves from coalescing binaries at cosmological distances could provide accurate measurements of the Universe's Hubble constant and mean density (Schutz 1986; Cutler et al. 1993; Chernoff and Finn 1993). Most importantly for our present discussion, the final stage of the coalescence, when the two neutron stars merge hydrodynamically, produces a burst of gravitational radiation whose characteristics probe directly the interior structure of a neutron star. The detection of this final burst of gravitational waves by LIGO would provide strong constraints on the interior structure of a neutron star and the nuclear equation of state at high densities (cf. RS1, RS2).

In addition, quite apart from their importance for LIGO, coalescing neutron-star binaries are also at the basis of numerous models of $\gamma$-ray bursters (Eichler et al. 1989; Meszaros and Rees 1992; Narayan, Paczyński, and Piran 1992; Davies et al. 1994; Mochkovich et al. 1995).

#### 1.1.2.2  Coalescing White-Dwarf Binaries

Double white-dwarf systems are now generally thought to be the progenitors of Type Ia supernovae (Iben and Tutukov 1984; Webbink 1984; Paczyński 1985; Yungelson et al. 1994). To produce a supernova, the total mass of the system must be above the Chandrasekhar mass. Given evolutionary considerations, this requires two C-O or O-Ne-Mg white dwarfs. Yungelson et al. (1994) show that the expected merger rate for close pairs of white dwarfs with total mass exceeding the Chandrasekhar mass is consistent with the rate of Type Ia supernovae deduced from observations. Alternatively, a massive enough merger may collapse to form a rapidly rotating neutron star (Nomoto and Iben 1985; Colgate 1990). Chen and Leonard (1993) have discussed the possibility that most millisecond pulsars in globular clusters may have formed in this way. In some cases planets may form in the disk of material ejected during the coalescence and left in orbit around the central pulsar (Podsiadlowski, Pringle, and Rees 1991). Objects of planetary masses in orbit around a millisecond pulsar (PSR B1257+12) have indeed been detected (Wolszczan 1994). A merger of two highly magnetized white dwarfs might lead to the formation of a neutron star with extremely high magnetic field, and this scenario has also been proposed as a source of gamma-ray bursts (Usov 1992).

Close white-dwarf binaries are expected to be extremely abundant in our Galaxy. Iben and Tutukov (1984, 1986) predict that $\sim 20\%$ of all binary stars produce close pairs of white dwarfs at the end of their stellar evolution. The most common systems should be those containing two low-mass helium white dwarfs. Their final coalescence can produce an object massive enough to start helium burning. Bailyn (1993) suggests that extreme horizontal branch stars in globular clusters may be such helium-burning stars formed by the coalescence of two white dwarfs. Paczyński (1990) has proposed that the peculiar X-ray pulsar 1E 2259+586 may be the product of a recent white-dwarf merger. Planets in orbit around a massive white dwarf may also form following a merger (Livio, Pringle, and Saffer 1992).

Coalescing white-dwarf binaries are also promising sources of low-frequency gravitational waves that should be easily detectable by future space-based interferometers such as the proposed LISA (Laser Interferometer Space Antenna, see Jafry, Cornelisse and Reinhard 1994; Evans, Iben, and Smarr 1987).

#### 1.1.2.3  Coalescing Main-Sequence Star Binaries

Two main-sequence stars in the process of merging (or about to merge) are directly observed as W Ursae Majoris contact systems (see Rucinski 1992a for a recent review). It is estimated that perhaps as much as 1 in every 150 stars in the Galaxy is a contact binary of the this type, containing two main-sequence stars embedded in a thin common envelope of gas (Rucinski 1994). The structure and evolution of these systems remains a major unsolved problem in stellar astrophysics. After heated debates between the proponents of various theoretical models in the late seventies (e.g., Shu 1980 for a brief account) the field went dormant, with a number of fundamental theoretical issues remaining unresolved. On the observational side, however, contact binaries continue to be discovered in large numbers and the quality of the data is improving steadily. It is likely that the rate of new detections will continue to increase in the near future. In particular, an important by-product of the recent searches for gravitational microlensing by MACHOS will be a large catalog of eclipsing binaries detected under very uniform conditions (Udalski et al. 1994). Interest in W UMa binaries was revived recently with the discovery of large numbers of new ones among blue stragglers in open and globular clusters (Kaluzny & Shara 1988; Kaluzny 1990; Mateo et al. 1990; Yan & Mateo 1994). It appears likely that at least some of the blue stragglers are formed by the merging of contact binaries. Rasio (1993, 1995) has pointed out the importance of instabilities in understanding the merging process.

### 1.1.2.4 Common Envelope Systems

All the coalescing binary systems mentioned above contain two stars of the same type. Instabilities can of course also affect systems containing two stars of different types, e.g., a main-sequence or giant star with a compact companion. Orbital decay in high-mass X-ray binaries (containing a massive main-sequence or giant star with an accreting neutron star companion) can be driven by a secular instability of the type discussed here. Examples of systems where rapid orbital decay has been directly observed include Cen X-3, where the orbital period is decreasing on a timescale $t_{\rm decay} \simeq 5 \times 10^5$ yr (Kelley et al. 1983), and SMC X-1, where $t_{\rm decay} \simeq 3 \times 10^5$ yr (Levine et al. 1993). In both cases it is thought that the unstable orbital decay is driven by tidal dissipation (Levine et al. 1991, 1993). In contrast, the orbital period of Her X-1 is changing on a much longer timescale $\simeq 10^8$ yr, consistent with the result that this system is most likely in a stable, synchronized configuration (Levine et al. 1991).

As the orbit of an unstable system is decaying rapidly, the primary will soon find itself overfilling its Roche lobe substantially, and mass transfer on a dynamical timescale will ensue. Since the mass transfer is from the primary to a much less massive secondary, the mass transfer rate will increase catastrophically on a dynamical timescale and the neutron star will rapidly spiral into the primary. The system then goes through a common envelope phase, with the neutron star and the compact helium core of the primary orbiting each other inside a common envelope of gas (Rasio and Livio 1995; Terman, Taam, and Hernquist 1995; see also the article by Livio in this volume). If the envelope is completely ejected at the end of the spiral-in phase, a new, much more compact binary system will have formed (see Verbunt 1993 for a discussion of various spiral-in scenarios). If the energy released during the common envelope phase is not sufficient to eject the envelope, then a massive Thorne-Zytkow object may be formed (Podsiadlowski, Cannon, and Rees 1995).

## 1.2  Dynamical Instabilities

### 1.2.1  Physical Mechanism

When the two components of a binary system are sufficiently close, the strong tidal interaction between the two stars can make the effective two-body potential governing their orbital motion deviate quite significantly from a Keplerian $1/r$ potential. Since the effective potential becomes *steeper* than $1/r$, a circular orbit can become *dynamically unstable* (see, e.g., Goldstein 1980, Chap. 2). A familiar example of this type of instability arises from general relativity, which can make circular orbits become unstable sufficiently close to a black hole (see, e.g., Shapiro and Teukolsky 1983, Chap. 12). Here, however, it is the purely Newtonian tidal interaction that causes the instability.

The existence of this dynamical instability in close binary configurations containing an *incompressible fluid* has been known for a long time (Chandrasekhar 1969, 1975; Tassoul 1975). However, the recent numerical work of RS and the analytical study of LRS have shown for the first time that dynamical instabilities continue to exist quite far into the compressible regime. For binaries containing polytropic components, it was shown that dynamical instabilities can exist for polytropic indices in the entire range $0 < n \lesssim 3/2$ (or, equivalently, for a fluid with adiabatic exponent in the range $5/3 \lesssim \gamma < \infty$; see especially RS3). In addition, the physical nature of the instability was elucidated in LRS by the use of an energy variational principle, which is conceptually and mathematically much simpler than the tensor virial method used previously.

The dynamical instability corresponds to an eigenmode of the system in which the predominant variation is that of the separation between the two components. One can think of this mode as describing small epicyclic oscillations of a close binary system, with the square of the epicyclic frequency $\omega_r^2$ becoming negative for an unstable system. However, in the calculation of $\omega_r^2$, it is essential to treat self-consistently the hydrodynamic response of the two stars themselves. This means not only including the self-gravity of the fluid in three dimensions, but also allowing for departures from a synchronized state. Indeed, even if the perturbed equilibrium configuration is assumed to be synchronized (i.e., with the entire mass of fluid rotating rigidly at the same angular frequency), synchronization is not maintained during dynamical perturbations. Instead, during dynamical oscillations, the *fluid circulation* is conserved. In the analytical treatment of LRS, a simple model of a close binary system is used in conjunction with an energy variational principle to determine both the (approximate) equilibrium solutions and to examine their stability. The equilibrium solutions are computed by finding stationary points of the energy function. Stability requires that the equilibrium solution correspond to a true minimum of the energy. The simple model must have enough parameters to allow for a self-consistent treatment of dynamical perturbations. LRS use compressible ellipsoids (basically the compressible generalizations of the classical Riemann ellipsoids; cf. Chandrasekhar 1969) to represent the two stars, and introduce explicitly the fluid circulation in the energy function.

A trivial example will serve to illustrate the energy method used by LRS.

Consider a test particle moving on a circular orbit around a Schwarzschild black hole of mass $M$. The total energy $E$ (the conserved "energy at infinity") describing circular motion of given angular momentum $J$ is given by (here in units where $G = c = 1$)

$$E(r; J) = m \left(1 - \frac{2M}{r}\right)^{1/2} \left(1 + \frac{J^2}{m^2 r^2}\right)^{1/2}, \quad (1.1)$$

where $m$ is the rest mass of the particle and $r$ is the Schwarzschild radial coordinate (see, e.g., Shapiro and Teukolsky 1983, Chap. 12). One can determine the "equilibrium solution" (i.e., the correct value of $r$ for a circular orbit of angular momentum $J$) by extremizing $E(r; J)$ at constant $J$. Eliminating $J$, we then get the following expression for the *equilibrium energy*,

$$E_{\rm eq}(r) = \frac{m(r - 2M)}{[r(r - 3M)]^{1/2}}. \quad (1.2)$$

It is easy to verify that $E_{\rm eq}(r)$ has a minimum at $r = 6M$, where the circular orbit becomes dynamically unstable. Equivalently, we could compute the second derivative of $E(r; J)$ and show that the equilibrium solution corresponds to a minimum of $E(r; J)$ for $r > 6M$ and a maximum of $E(r; J)$ for $r < 6M$. However, in more complicated systems, finding the minimum of $E_{\rm eq}$ is usually much simpler.

In fig. 1.1, we illustrate this technique for a simple model of a binary system containing two identical stars made of incompressible fluid (LRS1). The energy function in this case is written $E(r, \lambda_i; J, C)$, where the $\lambda_i$ are variables describing the internal degrees of freedom of the two stars, and $C$ is the conserved fluid circulation. Equilibrium solutions are found by solving the system of equations $\partial E/\partial r = 0$, $\partial E/\partial \lambda_i = 0$, where all partial derivatives are taken at constant $J$ and $C$. In practice, we construct a sequence of equilibrium configurations (in this case, synchronized configurations with varying separation $r$) and we determine the onset of dynamical instability along the sequence by locating the point where the equilibrium energy at constant $C$ is minimum. Note that along a sequence of synchronized equilibrium configurations the circulation $C$ is *not* constant, and the minimum energy along that sequence does *not* correspond to the onset of dynamical instability. Instead, the minimum of $E_{\rm eq}$ along the synchronized sequence marks the onset of *secular instability* (see sec. 1.3).

### 1.2.2 Application to Coalescing Neutron Star Binaries

An important application of dynamical instabilities has been to the study of gravitational wave emission from coalescing binary neutron stars (RS1, RS2). Indeed, for dynamical instabilities to play a significant role, the fluid in the system must be fairly incompressible. The fluid in a neutron star does have a rather stiff equation of state, with an adiabatic exponent most likely in the range $2 \lesssim \gamma \lesssim 3$.

Recent calculations of the gravitational radiation waveforms from coalescing neutron star binaries have focused on the signal emitted during the last few thousand orbits, as the orbit decays slowly due to the gravitational wave emission itself and the frequency of the waves sweeps upward from about 10 Hz to 1000 Hz.

**Figure 1.1**: Total equilibrium energy $E_{\rm eq}$ as a function of binary separation $r$ for a system containing two identical stars made of incompressible fluid. Here the separation is given in units of the (unperturbed) stellar radius $R_o$ and we have defined $\bar{E}_{\rm eq} = E_{\rm eq}/(GM^2/R_o)$, where $M$ is the mass of one star. The solid line corresponds to synchronized (rigidly rotating) equilibrium configurations, while the dashed line is for a sequence of equilibrium models with constant fluid circulation (which are nonsynchronized, except at the point of intersection with the solid line). The dynamical stability limit for synchronized systems corresponds to the point where the equilibrium energy along the intersecting constant-circulation sequence is minimum (square dot). The point of minimum equilibrium energy along the synchronized sequence itself corresponds to the secular stability limit (round dot).

The waveforms in this regime can be calculated fairly accurately by performing post-Newtonian expansions of the equations of motion for two *point masses* (see, e.g., Lincoln and Will 1990). However, at the end of the inspiral, the binary separation becomes comparable to the stellar radii and hydrodynamic effects become important, changing the character of the waveforms. In this terminal phase of the coalescence, the waveforms should contain information not just about the effects of relativity, but also about the radius and internal structure of a neutron star. Since the masses (and perhaps even the spins) of the two stars can in principle be determined accurately from the lower-frequency inspiral waveform, the measurement of a single quantity such as the peak amplitude or frequency of the signal could suffice to determine the neutron star radius and place severe constraints on nuclear equations of state (Cutler et al. 1992). Such a measurement requires theoretical knowledge about all relevant hydrodynamic processes.

Two regimes can be distinguished in which different hydrodynamic processes take place. The first regime corresponds to the $\sim 10$ or so orbits preceding the moment when the surfaces of the two stars first come into contact. In this regime, the two stars are still approaching each other in a quasi-static manner, but the tidal effects are becoming very large. The second regime corresponds to the subsequent merging of the two stars into a single object. This involves very large departures from hydrostatic equilibrium, including mass shedding and shocks, and can be studied only by means of fully three-dimensional hydrodynamic computations. Such three-dimensional computations have been attempted only recently. Nakamura and Oohara (1991 and references therein) and Ruffert, Janka, and Schäfer (1996) use traditional finite-difference techniques of numerical hydrodynamics, while RS, Davies et al. (1994), and Zhuge, Centrella, and McMillan (1995) use the smoothed particle hydrodynamics (SPH) method (see, e.g., Monaghan 1992).

In the first regime, the evolution of the system can still be described fairly accurately by a sequence of near-equilibrium fluid configurations. Such a description has been adopted in the recent work by LRS (see especially LRS2 and LRS3). An important result found in LRS3 is that equilibrium configurations for close binaries containing neutron stars with stiff equations of state ($\gamma \gtrsim 2$) always become dynamically unstable prior to contact. Thus the final merging of the two neutron stars is always driven by the (essentially Newtonian) hydrodynamic instability discussed in sec. 1.2.1, rather than being driven by the gravitational radiation reaction. The dynamical instability was indeed identified in RS1 and RS2, where the evolution of equilibrium configurations containing two identical polytropes with stiff equations of state ($2 \leq \gamma \leq 10$) was studied numerically using SPH. It was found that when the separation between the two stars becomes less than about three times the stellar radius (cf. fig. 1.1), the orbit is unstable and the stars merge in just a few orbital periods. For larger binary separations, the system could be evolved dynamically for many orbital periods without showing any sign of orbital evolution.

The details of the gravitational radiation waveforms emitted during the final coalescence depend sensitively on the neutron star equation of state. The equation of state of dense nuclear matter is not known. However, most detailed many-body calculations suggest that it is relatively stiff. For example, the Bethe-Johnson equa-tion of state is essentially a polytrope with $\gamma \simeq 2.5$ (Shapiro and Teukolsky 1983). Likewise, the more recent work of Wiringa, Fiks, & Fabronici (1988) predicts an effective $\gamma \gtrsim 2.3$ for all neutron star masses in the range $1\,M_\odot \lesssim M \lesssim 2\,M_\odot$ (see LRS3). Various nuclear phenomena such as a transition to a kaon condensate (cf. Brown and Bethe 1994) can soften this, but the importance of these phenomena is still not resolved. Because of this uncertainty, a simple polytropic equation of state was used in the numerical calculations of RS. The main results of RS can be summarized as follows. For $\gamma \lesssim 2.3$, RS find that the merger leads to the formation of a single, axisymmetric object. Since a stationary, axisymmetric configuration has no time-varying quadrupole, it does not radiate gravitational waves. As a result, the amplitude of the waves shuts off abruptly during the final merger (in just a few cycles, i.e., in a time $\sim 1\,\mathrm{ms}$). The position (e.g., in terms of frequency) of this sharp "edge" at the end of the signal marks the onset of dynamical instability in the system and is a direct measure of the radius of a neutron star. For $\gamma \gtrsim 2.3$, the final merged configuration is triaxial, and therefore continues to radiate gravitational waves. Thus both the position and the amplitude of the sudden decrease in the gravitational wave emission that accompanies the final merger can be used to constrain the equation of state.

## 1.3 Secular Instabilities

### 1.3.1 Physical Mechanism

Before a close binary system becomes dynamically unstable, and even if it never does, another type of global instability can occur. It has been referred to by various names, such as secular instability (LRS), tidal instability (Counselman 1973; Hut 1980), Darwin instability (Levine et al. 1993), or gravo-gyro instability (Hachisu and Eriguchi 1984). Its physical origin is very easy to understand. Consider a sequence of equilibrium binary configurations with decreasing binary separation $r$. In general, there exists a *minimum* value of the total angular momentum along this sequence. This is simply because the spin angular momentum, which *increases* as $r$ decreases, can become comparable to the orbital angular momentum for sufficiently small $r$. Now imagine a binary system undergoing slow orbital decay due to some angular momentum loss mechanism (such as magnetic breaking or the emission of gravitational waves). Assume that the tidal synchronization timescale is initially short compared to the angular momentum loss timescale. The evolution of the system will then proceed along the synchronized equilibrium sequence and the separation decreases at a rate given by $\dot{r} = \dot{J}/(dJ/dr)$, where $\dot{J}$ is the angular momentum loss rate and $dJ/dr$ is calculated along the equilibrium sequence. However, something must clearly go wrong in this picture near the minimum of $J$, since $dJ/dr \to 0$ there, giving $\dot{r} \to \infty$. Of course, in reality, when the system approaches the minimum of $J$, the orbital decay will be accelerated until it proceeds on a timescale so short that the tidal synchronization can no longer be maintained. Beyond that point, tidal dissipation will actually drive the system *out* of synchronization and cause rapid orbital decay (on the tidal dissipation timescale) as angular momentum is continually transferred from the orbit to the spin.

#### 1.3.1.1 Elementary Analytic Treatment

In contrast to the dynamical instability discussed in sec. 1.2, the secular instability can be understood on the basis of an elementary analytic model in which the stars are treated simply as spinning spheres. However, because the tidal effects are neglected in this model, it is accurate only for systems with well-separated components and rather extreme mass ratios. Indeed this model was originally developed for planet-satellite systems (see Counselman 1973).

Hut (1980, 1981, 1982) has provided a very elegant and exhaustive treatment of the problem, including a general classification of all possible types of tidal evolutions within this model (Hut 1981), and an extension to systems with highly eccentric orbits (Hut 1982). Here we merely summarize the results of his basic stability analysis (Hut 1980). Consider a binary system containing two spherical stars in a general elliptic Keplerian orbit of semimajor axis $a$ and eccentricity $e$. The total energy of the system may be written

$$E = -G\frac{M_1 M_2}{2a} + \frac{1}{2} I_1 \left|\mathbf{\Omega}_1\right|^2 + \frac{1}{2} I_2 \left|\mathbf{\Omega}_2\right|^2, \tag{1.3}$$

where $I_1$, $I_2$ are the moments of inertia and $\mathbf{\Omega}_1$, $\mathbf{\Omega}_2$ the spin vectors of the two stars. The total angular momentum is

$$\mathbf{J} = \mathbf{h} + I_1 \mathbf{\Omega}_1 + I_2 \mathbf{\Omega}_2, \tag{1.4}$$

where

$$h^2 = \frac{GM_1^2 M_2^2}{M_1 + M_2} a(1 - e^2). \tag{1.5}$$

The vectors $\mathbf{\Omega}_1$ and $\mathbf{\Omega}_2$ can have arbitrary directions, and their magnitude is not necessarily equal to the orbital angular velocity $\Omega_{\text{orb}} = [G(M_1 + M_2)/a^3]^{1/2}$. The orbital inclination $i$, defined as the angle between $\mathbf{h}$ and $\mathbf{J}$, is in general nonzero.

There are 9 relevant parameters in the model: $a$, $e$, $i$, and the components of $\mathbf{\Omega}_1$ and $\mathbf{\Omega}_2$. The equilibrium configurations for the system are found by determining stationary points of the energy $E$ in the six-dimensional subspace determined by the constraint of constant total angular momentum $\mathbf{J}$. Stability of an equilibrium configuration is then investigated by checking whether it corresponds to a true minimum of the energy. Hut (1980) does this by calculating explicitly all the eigenvalues of the matrix of second derivatives of $E$ (the Hessian). His main results can be summarized as follows:

1. All equilibrium configurations are circular ($e = 0$), coplanar ($i = 0$ and $\mathbf{h}$, $\mathbf{\Omega}_1$, $\mathbf{\Omega}_2$ all parallel), and synchronized ($\Omega_1 = \Omega_2 = \Omega_{\text{orb}}$).

2. Equilibrium solutions exist only if the total angular momentum $J$ is larger than a critical value

$$J_{\min} = 4 \left[ \frac{G^2}{27} \frac{M_1^3 M_2^3}{M_1 + M_2} (I_1 + I_2) \right]^{1/4}. \tag{1.6}$$

   Two solutions exist if $J > J_{\min}$, one if $J = J_{\min}$, zero if $J < J_{\min}$.

3. For $J > J_{\min}$, one of the two equilibrium solutions is secularly stable, the other is secularly unstable. The unstable solution is the one corresponding to the smaller separation $a$. Along a sequence of equilibrium configurations with decreasing $a$, the onset of instability corresponds to the point where the orbital angular momentum is equal to three times the total spin angular momentum, i.e.,

$$h = 3(I_1 + I_2)\Omega_{\rm orb} \qquad \text{(onset of instability)}. \tag{1.7}$$

   Equilibrium configurations with $h > 3(I_1 + I_2)\Omega_{\rm orb}$ are stable, those with $h < 3(I_1 + I_2)\Omega_{\rm orb}$ are unstable.

### 1.3.1.2 Secular Instabilities in Close Binary systems

For close binaries, the tidal deformations of the stars can become important and a more sophisticated analysis is required. Here again, we consider the model of LRS, where the stars are represented by compressible ellipsoids rather than spheres. This model provides much more accurate results for very close binaries, especially in the case of a rather incompressible fluid (cf. RS2). However, as shown below, the results are qualitatively similar to those of Hut (1980).

Since viscous dissipation conserves the total angular momentum of the system (but of course makes the total energy decrease), it is useful to consider sequences of equilibrium configurations with constant angular momentum (LRS4). In fig. 1.2 we show examples of constant−$J$ equilibrium curves for a system containing two identical $n = 1.5$ polytropes. Three values of $J$ are considered. We also show the synchronized sequence for comparison. All curves terminate at the point where the two stars come into contact. Note that the minima of $E$ and $J$ along the synchronized sequence coincide. This is a result of the general property that $dE = \Omega\, dJ$ along any sequence of uniformly rotating fluid equilibria (LRS1). We see that there exists a critical value $J = J_{min}$, equal to the minimum of $J$ along the synchronized sequence, above which a constant−$J$ sequence intersects the synchronized sequence. Moreover, the intersection points coincide exactly with an energy extremum along the constant−$J$ sequence. For the case shown in fig. 1.2a, the intersection point (point B) lies on the secularly stable branch of the synchronized sequence. Indeed, we see that this is the point of minimum energy along the corresponding constant−$J$ sequence, i.e., among all those configurations with the same $J$, the synchronized configuration is the one that has the lowest energy.

Note that if we consider a value of $J$ just slightly greater than $J_{min}$, the constant−$J$ curve can intersect the synchronized curve twice, once on the secularly stable branch and once on the unstable branch (fig. 1.2b). Both intersection points correspond to a local energy extremum along the constant-$J$ sequence. The intersection with the stable branch is a minimum (point H in fig. 1.2b), whereas the intersection with the unstable branch is a maximum (point I in fig. 1.2b). This explains very clearly the physical nature of the secular instability. Viscosity will drive a secularly unstable equilibrium configuration *away* from synchronization at first. As a result, the orbit can either expand (along IH) as the system is driven towards a lower-energy, stable synchronized state, or it can decay (along IJ) as the stars are driven to coalescence.

For the limiting case where $J = J_{min}$, the constant$-J$ sequence passes through the secular stability limit point along the synchronized sequence (point C in fig. 1.2a). In this case, the intersection is a stationary point of the equilibrium energy curve for the constant$-J$ sequence. Note that the first secularly unstable synchronized configuration at point C will always be driven to coalescence by viscous dissipation (in contrast to unstable configurations beyond C, for which the orbit can evolve either way). When $J < J_{min}$, the constant$-J$ sequence never intersects the synchronized sequence, and the energy decreases monotonically as $r$ decreases. Therefore, configurations with $J < J_{min}$ can never reach synchronization, and are always driven to coalescence by viscous dissipation.

The orbital evolution of an initially non-synchronized binary system depends critically on the total amount of angular momentum $J$ in the system. When $J < J_{min}$, as the the binary loses energy due to viscous dissipation, it simply slides down the constant$-J$ curve (e.g., along FG in fig. 1.2a). When $J > J_{min}$, the binary first evolves toward a (stable) synchronized configuration. If the star initially spins faster than the synchronized rate (point E in fig. 1.2a), the orbit expands as the system evolves toward synchronization (along EB). The Earth-Moon system is a well-known example of such an evolution. If the initial spin is slower than the synchronized rate (point A), then the separation decreases as the binary evolves toward synchronization (along AB).

### 1.3.2 Application to Contact Binaries

As a simple application of these ideas, consider the stability of contact main-sequence star (W UMa type) binaries (Rasio 1995). The smallest observed mass ratio for these systems is $q \simeq 0.075$ (for AW UMa; cf. Rucinski 1992b). We will show below that this observed minimum mass ratio can be explained naturally from stability considerations. Systems with mass ratios below that of AW UMa may not be observed simply because they are unstable and undergo rapid merging. The timescale for merging can be estimated from standard tidal dissipation theory and should be of order $t_{\mathrm{decay}} \sim 10^3 - 10^4$ yr for these systems. This is much shorter than the typical timescale of observed orbital period changes in W UMa binaries.

Because we are interested in systems with rather extreme mass ratios, a simple analysis assuming a spherical primary will be sufficient here (cf. sec. 1.3.1.1). Specifically, we consider a close binary in which the primary is filling its Roche lobe, and with a small mass ratio $q = M_2/M_1 \ll 1$. Assume also that the secondary is much smaller than the primary, so that we can neglect its spin angular momentum. We can then write the orbital angular momentum $h$ and the spin angular momentum $J_s$ as

$$h = \mu a^2 \Omega_{\mathrm{orb}}; \qquad J_{\mathrm{s}} = M_1 k_1^2 R_1^2 \Omega_{\mathrm{orb}}. \qquad (1.8)$$

Here $a$ is the binary separation, $\Omega_{\mathrm{orb}}$ is the orbital frequency, $k_1$ is the dimensionless gyration radius of the primary, and $\mu = M_1 q/(1+q)$ is the reduced mass.

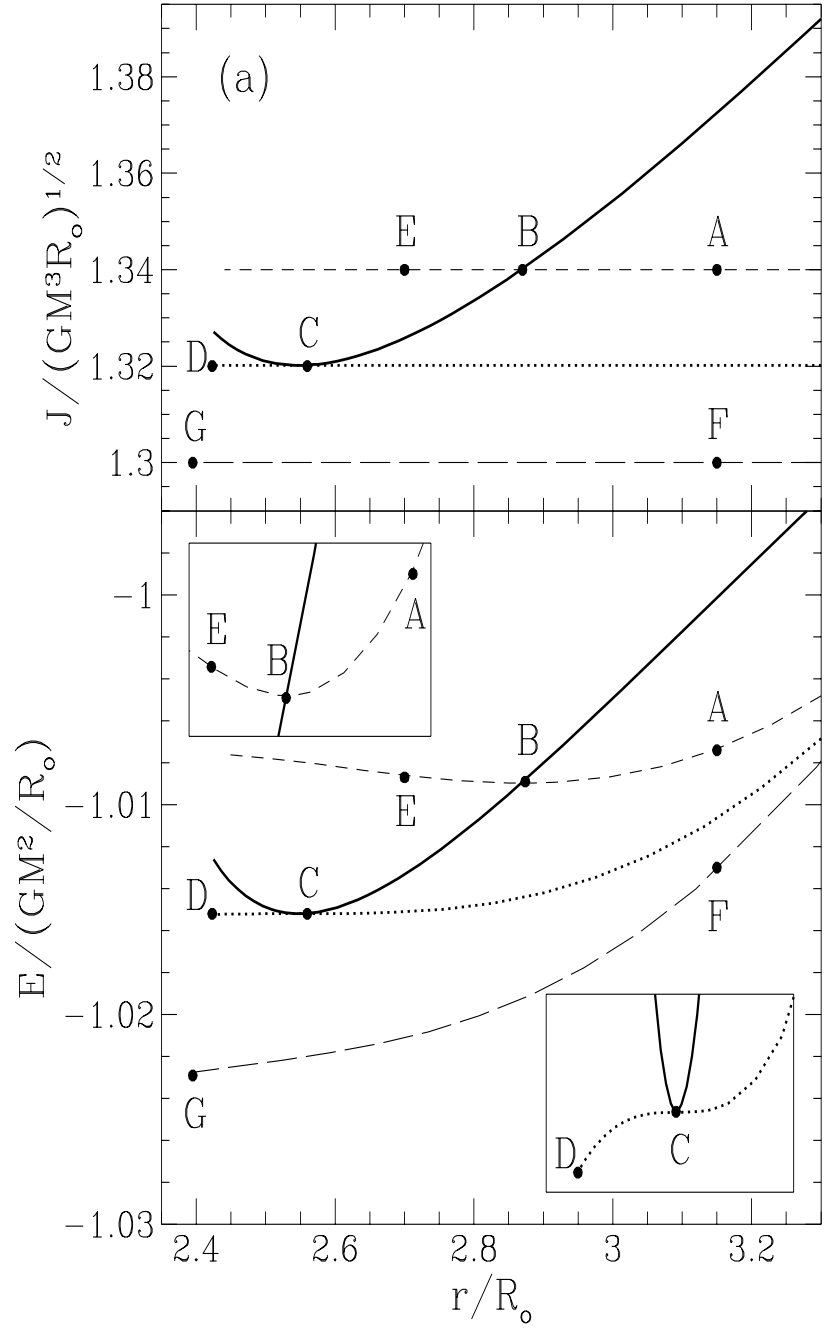

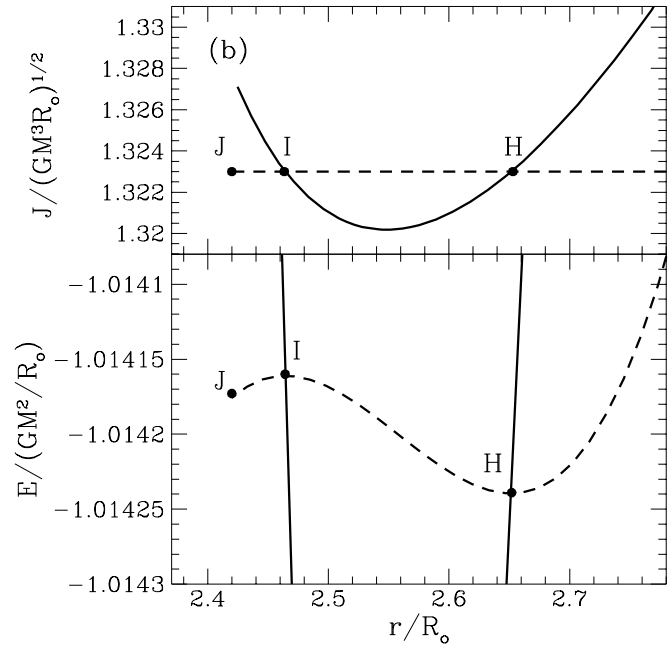

**Figure 1.2**: Total energy $E$ and total angular momentum $J$ as a function of binary separation $r$ for a system containing two identical $n = 1.5$ polytropes. Here the separation is given in units of the (unperturbed) stellar radius $R_o$. The solid lines correspond to synchronized (rigidly rotating) equilibrium configurations. The dotted and dashed lines correspond to sequences of equilibrium models with constant angular momentum (which are generally nonsynchronized). In (a), the regions around points B and C are magnified in the inserts. See text for discussion.

The onset of instability corresponds to $h = 3J_s$ (cf. eq. 1.7), or a critical separation $a_{\rm inst}$ given by

$$\frac{a_{\rm inst}}{R_1} = \left[\frac{3(1+q)}{q}\right]^{1/2} k_1. \qquad (1.9)$$

This must be compared to the separation $a_{\rm RL}$ at the Roche limit, for which we can write

$$\frac{a_{\rm RL}}{R_1} = \frac{a_{\rm RL}}{\bar{R}_{\rm L1}} = \frac{0.6 + q^{2/3}\ln(1+q^{-1/3})}{0.49}, \qquad (1.10)$$

using Eggleton's (1983) fitting formula for the volume mean radius $\bar{R}_{\rm L1}$ of the primary's Roche lobe.

Setting $a_{\rm inst} = a_{\rm RL}$ and solving (numerically) for $q$, we obtain the *minimum mass ratio for stability*, $q_{\rm min}$. A system with $q < q_{\rm min}$ cannot exist in a stable contact configuration. The results are plotted in fig. 1.3 as a function of $k_1$, the only parameter in this simple model. To within $\sim 20\%$, one has the remarkably simple approximate solution $q_{\rm min} \simeq k_1^2$.

The results obtained from eqs. (1.8)–(1.10) are only valid as long as $q_{\rm min} \ll 1$, and for a centrally condensed primary. In general, fully three-dimensional numerical calculations must be performed. This was done recently in RS3 for a system containing two fully convective MS stars (modelled as $n = 1.5$ polytropes with the polytropic constants adjusted so as to obtain a simple mass-radius relation with $R \propto M$). For such a system it is found that $q_{\rm min} \simeq 0.45$. As can be seen in fig. 1.3, the rigid sphere and Roche approximations fail badly in this case, underestimating $q_{\rm min}$ by about a factor 2. For binaries containing stars even less compressible than $n = 1.5$ polytropes, such as neutron stars, it is possible to find that the binary always becomes secularly unstable before reaching contact, for all mass ratios (RS2). However, for sufficiently compressible stars, such as a mostly radiative main-sequence star (well modeled by a polytrope with $n = 3$, $\Gamma_1 = 5/3$) numerical results indicate that eqs. (1.8)–(1.10) are an excellent approximation.

We can now examine the consequences of these results for contact binaries. To obtain $q_{\rm min} = 0.075$ (placing AW UMa just at the stability boundary), the primary must have $k_1^2 \simeq 0.06$. This is less than the value for an $n = 3$ polytrope, which has $k^2(n = 3) = 0.08$. The immediate implication is that the primary in AW UMa cannot have much of a convective envelope, and must be slightly evolved. The latter point is consistent with this system being classified observationally as an "A-type" system (primary eclipse is a transit of the secondary; see, e.g., Rucinski 1985). The former is in conflict with at least some simple models for the interior structure of contact binaries (Rucinski 1992a). The broader implication, of course, is that contact binaries with mass ratios below that of AW UMa may not be observed simply because they are unstable and undergo rapid merging.

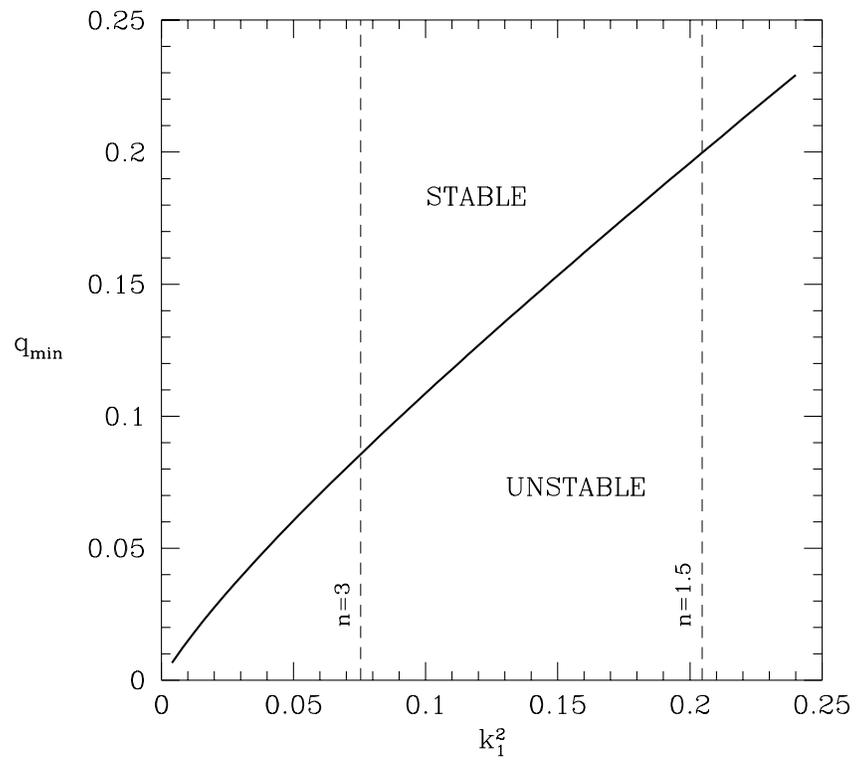

**Figure 1.3**: The minimum mass ratio for secular stability, $q_{\min}$, calculated for a spherical primary in the Roche approximation (solid line). The parameter $k_1^2$ is the dimensionless gyration radius of the primary (values for $n = 1.5$ and $n = 3$ polytropes are indicated by the vertical dashed lines).

*Acknowledgements* The author thanks M. Davies and R. Wijers for their hospitality at the Institute of Astronomy, and for their patience while this manuscript was being prepared. This work was supported in part by the NSF and NASA. Some of the results were obtained from computations performed at the Cornell Theory Center.

# References

Abramovici, A., et al.: 1992, *Science* **256**, 325
Alexander, M. E.: 1973, *Astrophys. Space Sci.* **23**, 459
Bailyn, C. D.: 1993, in S. Djorgovski and G. Meylan (eds.), *Structure and Dynamics of Globular Clusters*, pp 191–201, ASP Conf. Series Vol. 50, San Francisco
Bodenheimer, P., and Ostriker, J. P.: 1973, *Astrophys. J.* **180**, 159
Brown, G. E., and Bethe, H. A.: 1994, *Astrophys. J.* **423**, 659
Chandrasekhar, S.: 1969, *Ellipsoidal Figures of Equilibrium*, Yale Univ. Press, New Haven
Chandrasekhar, S.: 1975, *Astrophys. J.* **202**, 809
Chen, K, and Leonard, P. J. T.: 1993, *Astrophys. J. Letters* **411**, L75
Chernoff, D. F., and Finn, L. S.: 1993, *Astrophys. J. Lett.* **411**, L5
Colgate, S. A.: 1990, in S. E. Woosley (ed.) *Supernovae*, pp 585–592, Springer-Verlag, New York
Counselman, C. C.: 1973, *Astrophys. J.* **180**, 307
Cutler, C., et al.: 1993, *Phys. Rev. Lett.* **70**, 2984
Davies, M. B., Benz, W., Thielemann, F. K., and Piran, T.: 1994, *Astrophys. J.* **431** , 742
Deeter, J. E., Boynton, P. E., Miyamoto, S., Kitamoto, S., Nagase, F., and Kawai, N.: 1991, *Astrophys. J.* **383**, 324
Eggleton, P. P.: 1983, *Astrophys. J.* **268**, 368
Eichler, D., Livio, M., Piran, T., Schramm, D. N.: 1989, *Nature* **340**, 126
Evans, C. R., Iben, I., and Smarr, L.: 1987, *Astrophys. J.* **323**, 129
Finn, L. S., and Chernoff, D.: 1993, *Phys. Rev. D* **47**, 2198
Goldstein, H.: 1980, *Classical Mechanics* 2nd ed., Addison-Wesley, Reading
Hachisu, I., and Eriguchi, Y.: 1984, *Pub. Astron. Soc. Japan* **36**, 239
Hut, P.: 1980, *Astron. & Astrophys.* **92**, 167
Hut, P.: 1981, *Astron. & Astrophys.* **99**, 126
Hut, P.: 1982, *Astron. & Astrophys.* **110**, 37
Iben, I., and Tutukov, A. V.: 1984, *Astrophys. J. Suppl.* **54**, 335
Iben, I., and Tutukov, A. V.: 1986, *Astrophys. J.* **311**, 753
Jafry, Y. R., Cornelisse, J., and Reinhard, R.: 1994, *ESA Journal* **18**, 219
Kaluzny, J.: 1990, *Acta Astron.* **40**, 61
Kaluzny, J., and Shara, M. M.: 1988, *Astron. J.* **95**, 785
Kelley, R. L., Rappaport, S. A., Clark, G. W., and Petro, L. D.:1983
Lai, D., Rasio, F. A., and Shapiro, S. L.: 1993a, *Astrophys. J. Suppl.* **88**, 205 [LRS1]

Lai, D., Rasio, F. A., and Shapiro, S. L.: 1993b, *Astrophys. J. Letters* **406**, L63 [LRS2]
Lai, D., Rasio, F. A., and Shapiro, S. L.: 1994a, *Astrophys. J.* **420**, 811 [LRS3]
Lai, D., Rasio, F. A., and Shapiro, S. L.: 1994b, *Astrophys. J.* **423**, 344 [LRS4]
Lai, D., Rasio, F. A., and Shapiro, S. L.: 1994c, *Astrophys. J.* **437**, 742 [LRS5]
Levine, A., Rappaport, S., Putney, A., Corbet, R., and Nagase, F.: 1991, *Astrophys. J.* **381**, 101
Levine, A., Rappaport, S., Deeter, J. E., Boynton, P. E., and Nagase, F.: 1993, *Astrophys. J.* **410**, 328
Lincoln, W., and Will, C.: 1990, *Phys. Rev. D* **42**, 1123
Livio, M., Pringle, J. E., and Saffer, R. A.: 1992, *Mon. Not. RAS* **257**, 15P
Mateo, M., Harris, H. C., Nemec, J., and Olszewski, E. W.: 1990, *Astron. J.* **100**, 469
Meszaros, P., and Rees, M. J.: 1992, *Astrophys. J.* **397**, 570
Mochkovitch, R., Hernanz, M., Isern, J., Loiseau, S.: 1995, *Astron. Astrophys.* **293**, 803
Monaghan, J. J.: 1992, *Ann. Rev. Astron. & Astrophys.* **30**, 543
Nakamura, T., and Oohara, K.: 1991, *Prog. Theor. Phys.* **86**, 73
Narayan, R., Paczyński, B., and Piran, T.: 1992, *Astrophys. J. Letters* **395**, L83
Narayan, R., Piran, T., and Shemi, A.: 1991, *Astrophys. J. Lett.* **379**, L17
Nomoto, K.: 1987, in D. J. Helfand and J.-H. Huang (eds.), *IAU Symposium 125, Origin and Evolution of Neutron Stars*, pp 281–289, Reidel, Dordrecht
Nomoto, K., and Iben, I., Jr.: 1985, *Astrophys. J.* **297**, 531
Paczyński, B.: 1985, in D. Q. Lamb and J. Patterson (eds.) *Cataclysmic Variables and Low-mass X-ray Binaries*, pp 1–13, Reidel, Dordrecht
Paczyński, B.: 1990, *Astrophys. J. Lett.* **365**, L9
Phinney, E. S.: 1991, *Astrophys. J. Lett.* **380**, L17
Podsiadlowski, P., Pringle, J. E., and Rees, M. J.: 1991, *Nature* **352**, 783
Podsiadlowski, P., Cannon, R. C., and Rees, M. J.: 1995, *Mon. Not. RAS* **274**, 485
Rasio, F. A.: 1993, in R. E. Saffer (ed.), *Blue Stragglers*, pp 196–200, ASP Conf. Series Vol. 53, San Francisco
Rasio, F. A.: 1995 *Astrophys. J. Lett.* **444**, L41
Rasio, F. A., and Livio, M.: 1995, *Astrophys. J.*, submitted
Rasio, F. A., and Shapiro, S. L.: 1992, *Astrophys. J.* **401**, 226 [RS1]
Rasio, F. A., and Shapiro, S. L.: 1994, *Astrophys. J.* **432**, 242 [RS2]
Rasio, F. A., and Shapiro, S. L.: 1995, *Astrophys. J.* **438**, 887 [RS3]
Rucinski, S. M.: 1985, in J. E. Pringle and R. A. Wade (eds.) *Interacting Binary Stars* pp 85–102, Cambridge Univ. Press, Canbridge
Rucinski, S. M.: 1992a, in J. Sahade et al. (eds.), *The Realm of Interacting Binary Stars* pp 111-XX, Kluwer, Dordrecht
Rucinski, S. M.: 1992b, *Astron. J.* **104**, 1968
Rucinski, S. M.: 1994, *Pub. Astron. Soc. Pacific* **106**, 462
Ruffert, M., Janka, H.-T., Schäfer, G.: 1996, *Astron. & Astrophys.*, submitted
Schutz, B. F.: 1986, *Nature* **323**, 310
Shapiro, S. L. and Teukolsky, S. A.: 1983, *Black Holes, White Dwarfs, and Neutron*

   *Stars*, Wiley, New York
Shu, F. H.: 1980, in M. J. Plavec, D. M. Popper and R. K. Ulrich (eds.) *Close Binary Stars: Observations and Interpretation*, IAU Symp. 88, pp 477–486, Reidel), Dordrecht
Tassoul, M.: 1975, *Astrophys. J.* **202**, 803
Terman, J. L., Taam, R. E., and Hernquist, L.: 1995, *Astrophys. J.* **445**, 367
Udalski, A., Kubiak, M., Szymański, M., Kaluzny, J., Mateo, M., and Krzemiński, W.: 1994, *Acta Astron.* **44**, 317
Usov, V. V.: 1992, *Nature* **357**, 472
Verbunt, F.: 1993, *Ann. Rev. Astron. Astrophys.* **31**, 93
Webbink, R. F.: 1984, *Astrophys. J.* **277**, 355
Wiringa, R. B., Fiks, V., and Fabronici, A.: 1988, *Phys. Rev. C* **38**, 1010
Wolszczan, A.: 1994, *Science* **264**, 538
Yan, L., and Mateo, M.: 1994, *Astron. J.* **108**, 1810
Yungelson, L. R., Livio, M., Tutukov, A. V., and Saffer, R. A.: 1994, *Astrophys. J.* **420**, 336
Zahn, J.-P.: 1977, *Astron. & Astrophys.* **57**, 383
Zughe, X., Centrella, J. M., & McMillan, S. L. W.: 1995, *Phys. Rev. D*, in press